\documentclass[acmsmall,nonacm]{acmart}

\usepackage[yyyymmdd]{datetime}

\usepackage{tabularx}
\usepackage{booktabs}
\usepackage{array}
\usepackage{tabularx}
\usepackage{booktabs}  
\usepackage{siunitx}   

\usepackage{dcolumn} 
\newcolumntype{d}[1]{D{.}{.}{#1}}

\usepackage{subcaption}
\usepackage{xcolor}

\usepackage{colortbl}
\usepackage[normalem]{ulem}
\usepackage{xcolor}

\definecolor{DeleteGrey}{RGB}{120,120,120}
\definecolor{AddColor}{RGB}{0,110,255}

\newif\ifrevisions

\revisionsfalse  

\ifrevisions

    \newcommand{\add}[1]{\textcolor{AddColor}{#1}}
    \newcommand{\delete}[1]{\textcolor{AddColor}{\sout{#1}}}

\else

    \newcommand{\add}[1]{#1}
    \newcommand{\delete}[1]{}

\fi

\AtBeginDocument{%
  }

\begin{document}

\footnotesize{
\noindent
This is the author's accepted manuscript of an article
accepted for publication in Proceedings of the ACM on
Human-Computer Interaction (PACM-HCI). The final version of record is available at https://doi.org/10.1145/3821685.
\\
}

\title[Typing Behavior in Human-LLM Interaction]{\add{Typing Behavior in Human-LLM Interaction: Keystroke Dynamics Reveal Cognitive Effort During Prompting}}


\settopmatter{authorsperrow=3}

\author{Laura Schütz}
\orcid{0000-0002-5534-3903}
\affiliation{
  \institution{Technical University of Munich}
  \city{Munich}
  \country{Germany}
}
\email{laura.schuetz@tum.de}

\author{Yousri Cherif}
 \orcid{0009-0002-9370-2468}
\affiliation{
  \institution{LMU Munich}
  \city{Munich}
  \country{Germany}
}
\email{cherif.yousri@campus.lmu.de}

\author{Clara Sayffaerth}
\orcid{0009-0005-4880-8572}
\affiliation{
  \institution{LMU Munich}
  \city{Munich}
  \country{Germany}
}
\email{clara.sayffaerth@ifi.lmu.de}

\author{Thomas Weber}
\orcid{0000-0002-6894-605X}
\affiliation{%
 \institution{LMU Munich}
 \city{Munich}
 \country{Germany}
}
\email{thomas.weber@ifi.lmu.de}

\author{Francesco Chiossi}
\orcid{0000-0003-2987-7634}
\affiliation{
  \institution{LMU Munich}
  \city{Munich}
  \country{Germany}
}
\email{francesco.chiossi@ifi.lmu.de}

\renewcommand{\shortauthors}{Schütz et al.}

\begin{abstract}
As Large Language Models (LLMs) become increasingly integrated into daily routines, understanding how users interact with these systems is crucial for effective human–AI collaboration. This work investigates keystroke dynamics as a behavioral measure \delete{for evaluating user experience and perceived usefulness of LLM outputs}\add{of user mental effort and perceived output usefulness in human-LLM interaction}. We conducted a user study (N = 36) to examine how task difficulty (easy vs. hard) and device type (desktop vs. mobile) influence typing behavior and workload (NASA-TLX) during interactions. Our results indicate that hard tasks led to significantly more keystrokes, slower typing, increased pauses, and higher self-reported workload. Device type had weaker effects, with mobile use slightly reducing input length and typing speed. While keystrokes captured differences in cognitive effort, they did not predict perceived LLM output usefulness. These findings highlight the potential of keystroke dynamics as real-time indicators of cognitive effort during LLM prompting, while also showing their limitations in capturing perceived collaboration success.
\end{abstract}




\begin{teaserfigure}
 \includegraphics[width=\linewidth]{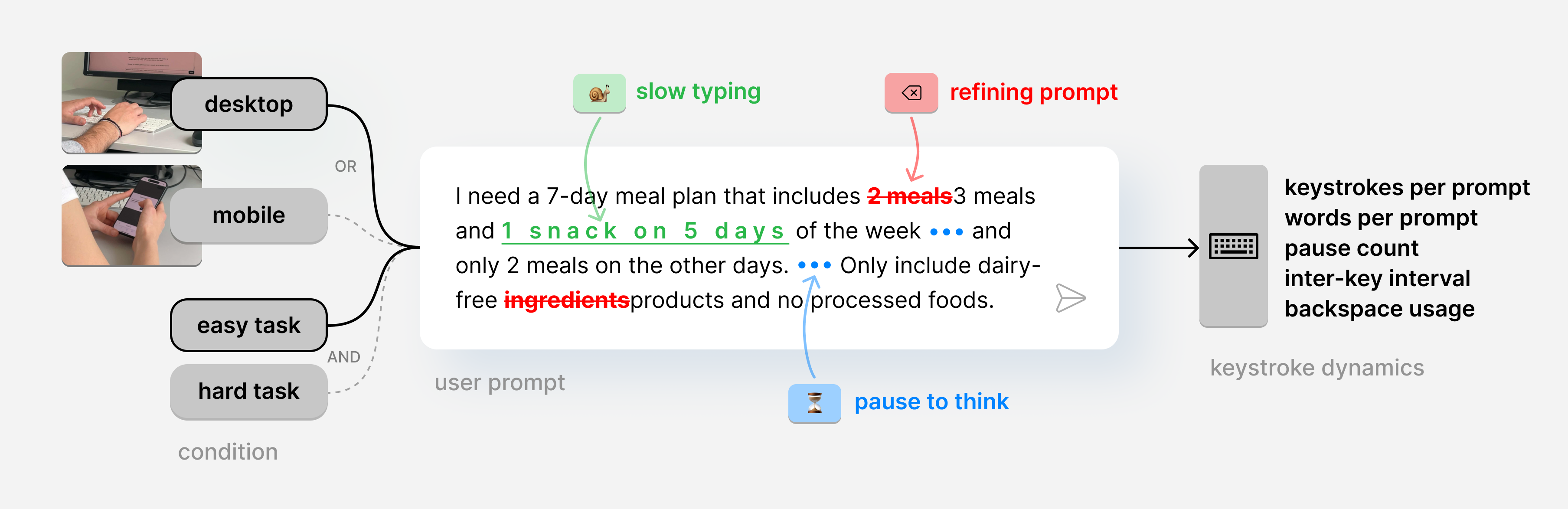}
\caption{\add{We investigated how device type (mobile versus desktop) and task difficulty (easy versus hard) shape typing behavior during human--LLM interaction. Keystroke dynamics reflected users' cognitive effort during LLM prompting across desktop and mobile devices, but did not reliably predict perceived output usefulness.}}
 \Description{This horizontal flow diagram illustrates the experimental setup for a study on human-LLM interaction, moving from left to right to show how conditions influence behavior and data collection. On the left, the "condition" section identifies the variables being tested: device type (desktop versus mobile) and task difficulty (easy versus hard). The center of the image features a "user prompt" box containing a meal plan request that is annotated with behavioral markers, such as a snail icon for slow typing, red strike-throughs for prompt refinement, and hourglass icons indicating pauses to think. An arrow connects this prompt to a final "keystroke dynamics" box on the right, which lists the specific metrics captured during the interaction, including keystrokes and words per prompt, pause count, inter-key interval, and backspace usage.}
 \label{fig:teaser}
\end{teaserfigure}

\maketitle

\section{Introduction}

Large language models (LLMs) are rapidly transforming how humans interact with technology. In just a short period, LLM-powered systems like ChatGPT have become one of the fastest adopted technologies in history~\cite{todd2025ai}, offering tremendous productivity benefits across professional writing~\cite{noy2023experimental}, software development~\cite{ziegler2022neural, weber2024significant}, testing~\cite{weisz2025examining}, and debugging~\cite{pearce2022examiningzeroshotvulnerabilityrepair}. Unlike traditional software interfaces, where users issue explicit commands, conversational chat interfaces enable users to collaborate with AI through natural language dialogue. This shift transforms users from command-givers into collaborators who iteratively refine their intent through multi-turn exchanges with an intelligent system. As LLM-powered tools become embedded in everyday workflows, understanding the behavioral dynamics of this collaboration, particularly during the prompting process itself, becomes critical for designing systems that can adapt to users' needs in real time.

However, existing evaluations of human-LLM interaction have largely relied on post-hoc methods, subjective user perception ratings~\cite{weisz2025examining}, or outcome-based accuracy metrics that require ground truth. While useful for summative assessment, these approaches cannot capture the temporal and behavioral dynamics of interaction as it unfolds. They provide no insight into how users experience collaboration during prompting and interacting with an LLM: whether they struggle to articulate intent, feel frustrated by inadequate responses, or invest excessive cognitive effort in refinement. 
The absence of such insight, in turn, restricts opportunities for adaptive system design that could respond to user state in real time.

Keystroke dynamics offer a promising alternative. As text-based input remains the primary interaction modality for most LLM systems, typing behavior is both ubiquitous and rich in information. Prior research has demonstrated that keystroke patterns, e.g., pauses, inter-key intervals, and correction behavior, can reveal mental fatigue~\cite{acien2022detection}, cognitive load~\cite{brizan2015utilizing}, and emotional state~\cite{nahin2014emotion, epp2011identifying}. Unlike physiological sensors or intrusive self-reports, keystrokes can be logged unobtrusively and continuously~\cite{shadman2025keystrokes, liu2026Sensing, chiossi2024evaluating}, making them particularly suitable for real-world deployment. However, little is known about how these signals manifest during LLM-assisted tasks, or whether they can effectively capture the quality of human-AI collaboration across different devices.

\add{We hypothesize that keystrokes during prompting might differ from those in existing chat-based interactions due to the cognitive processes at play during prompting, including deliberations on how to instruct the LLM to achieve the intended goal, and which output to expect based on the provided prompt~\cite{subramonyam2024gulf}. These cognitive steps of planning and anticipating interaction outcomes lead to iterative prompt refinement~\cite{gao2024taxonomy}, potentially resulting in more pauses and higher correction rates. 
Furthermore, we hypothesize that dissatisfaction with LLM output, a negative arousal state, is likely to produce measurable shifts in typing dynamics. Previous research has shown that high-arousal negative states significantly affect, e.g., inter-key latency, error rate, and deletion activity~\cite{epp2011identifying, lee2015influence, nahin2014emotion, kolakowska2015emotions}.
Thus, we investigated whether typing behavior during prompting can serve as a real-time signal of cognitive effort and perceived usefulness of LLM output.}

\add{To test these hypotheses, we conducted} a controlled study with 36 participants \add{to} examine how task difficulty (easy vs. hard) and device type (desktop vs. mobile) shape typing behavior during iterative prompt refinement. We developed a web-based platform that logs detailed keystroke data, including typing speed, inter-key intervals, pause patterns, and correction behavior, while participants collaborate with an LLM to generate meal plans with varying levels of constraint complexity. Alongside behavioral metrics, we collected self-reported measures of cognitive workload (NASA-TLX), perceived usefulness of AI outputs, and the difficulty of refining after each interaction turn. This multi-level analysis enables us to assess whether keystroke dynamics can serve as continuous, implicit indicators of both user effort and collaboration success.

Our findings reveal both the potential and boundaries of keystroke-based evaluation for human-AI collaboration. Typing behavior proved highly sensitive to task complexity: harder tasks elicited significantly more keystrokes, slower typing, increased pauses, and higher mental workload. Device type had weaker effects, with mobile users showing only modest reductions in speed and input length, underscoring the robustness of keystroke metrics across devices. Critically, however, we found no relationship between keystroke metrics and perceived usefulness of AI responses, thus suggesting that while keystroke dynamics capture user effort during collaboration, they do not capture judgments about collaboration success. These findings position keystroke dynamics as valuable real-time indicators of cognitive effort, while highlighting the need to combine behavioral signals with semantic evaluation for comprehensive assessment of human-\add{LLM} interaction.

\section{Related Work}

In the following, we highlight examples of human-AI collaboration to motivate the potential benefit of keystroke dynamics as a dimension in human-AI interaction. Following this, we will go into further detail on prior work investigating keystroke dynamics to model user cognition and behavior.

\subsection{Human-AI Collaboration}
With the widespread adoption of large language models and conversational agents, human-AI collaboration has become a growing area of interest in HCI. While AI promises to enhance productivity and decision-making, research reveals diverse challenges in achieving effective collaboration and in measuring it.
A systematic review by \citet{vaccaro2024when} found that, on average, human-AI systems performed significantly worse than the best individual agent, whether human or AI alone. In scenarios where AI systems outperformed humans, integrating human input reduced performance. However, when humans outperformed AI, hybrid teams could sometimes achieve performance gains through synergy. Critically, outcomes depended on task type, data characteristics, and AI system design, highlighting that interaction design plays a central role in collaboration success.

Traditional evaluation methods, i.e., outcome-based metrics and post-hoc subjective ratings, have proven insufficient for capturing the temporal dynamics of this collaboration. \citet{simkute2025ironies} identified four mechanisms of productivity loss during generative AI use: users shifting from active production to passive evaluation, workflows being restructured unhelpfully through prompting overhead, flow-disrupting interruptions from AI suggestions, and task-complexity polarization where AI simultaneously simplifies easy tasks but complicates harder ones by adding monitoring demands. To mitigate these issues, the authors proposed carefully timing system interventions to preserve user flow states, yet this requires real-time awareness of user cognitive state, something post-hoc surveys cannot provide.

Evidence of process-outcome dissociations further identifies this measurement gap. 
\citet{qian2024take} observed that 76 software engineers spent more time on programming tasks when using conversational AI, yet perceived themselves as more productive, showing a divergence between perceived and actual efficiency that post-hoc subjective ratings alone cannot explain. This dissociation suggests that users' retrospective judgments may not accurately reflect the cognitive costs incurred during interaction.
\delete{Weber et al. \cite{weber2024significant} found significant benefits of leveraging AI for programming. Comparing different interaction modalities for AI-supported programming, their work shows that the AI interface's interaction design affects users' typing behavior.}

To reduce this gap, providing \add{AI systems} with information about the user during interaction could help adapt the AI's output to the user's needs, leading to better collaboration overall.
\citet{hemmer2023human}, for example, propose an approach where tasks are automatically delegated between human and AI to ensure that the most suitable actor performs the task. For this to be effective, the decision logic does not only need information about the general capabilities of humans and AI but would also benefit from understanding the current state of humans, e.g., their cognitive load, emotional state, etc. Again, keystroke dynamics is one avenue for extracting this information in an unobtrusive way.

\subsection{Keystroke Dynamics} 
Keystroke dynamics, such as the rhythm, timing, and patterns of keyboard input\delete{.} offer a behavioral window into users' cognitive and affective states ~\cite{yang2025identify}. 
Unlike physiological sensors that require specialized equipment, keystroke data are inherently present in text-based interaction and can be logged unobtrusively using standard hardware~\cite{shadman2025keystrokes}. Common metrics include timing features (inter-key intervals, pause duration, words per minute), error-related behavior (backspace frequency, correction patterns), and structural features (input length, sentence complexity). These metrics capture individual differences in typing style while remaining sensitive to momentary changes in user state~\cite{yang2025identify}.

\subsubsection{Keystrokes and Cognitive Load} 
A substantial body of work has established that keystroke patterns reflect cognitive effort during text production. \citet{brizan2015utilizing} demonstrated that typing speed, pause frequency, and linguistic complexity could reliably distinguish between low and high cognitive demand tasks. \citet{nie2022time} showed that cognitive load impacts keystroke timing in quantifiable ways, with increased mental demand producing measurable changes in typing rhythm that can provide real-time feedback on user state~\cite{vizer2009detecting}. 

However, the specific metrics that prove diagnostic vary across contexts. \citet{conijn2019understanding} compared keystroke behavior during email writing versus essay writing, finding that word count, revision frequency, and total time correlated with task complexity, while inter-key intervals did not reliably indicate cognitive load. This suggested that metric validity may depend on task characteristics. In contrast, \citet{oliveira2020writing} observed that essay writing under higher cognitive load produced increased inter-word intervals but fewer revisions per minute, indicating users adopted a more deliberate, less iterative composition strategy. \citet{likens2017keystrokes} found that more structured, consistent typing rhythms predicted higher essay quality, suggesting fluency-related metrics may capture both momentary cognitive state and domain expertise. In programming contexts, \citet{shrestha2022pausing} found that pause frequency negatively correlated with code performance, supporting the interpretation that hesitations reflect uncertainty or difficulty. 

Together, these findings demonstrate that keystroke dynamics are sensitive to cognitive effort, but the relationship between specific metrics and performance outcomes appears task-dependent. Beyond cognitive load, keystroke dynamics have also been shown to reflect mental fatigue, with variations in typing patterns serving as passive indicators of declining user wellbeing during extended computer use~\cite{acien2022detection}.

\subsubsection{Affective States and Behavioral Context}

Keystroke patterns capture not only cognitive effort but also emotional states. \citet{epp2011identifying} demonstrated that typing dynamics could distinguish among confidence, hesitance, nervousness, relaxation, and sadness with reasonable accuracy. Similar detection capabilities have been reported for other affective states~\cite{qi2021emotion, marrone2022identifying}, suggesting that keystroke dynamics reflect both how hard users are working and how they feel while doing so.
\add{The same is true for typing behavior on mobile devices. Previous works in HCI have predicted user emotions from touch inputs like keystrokes and swiping motions using machine learning classification methods on keystroke logs~\cite{ghosh2017tapsense, GHOSH2019emotion} or interaction heat maps~\cite{wampfler2020affective, wampfler2022affective}.}

Recent HCI research has further revealed that keystroke behavior is shaped by contextual and environmental factors. Large-scale in-situ studies such as ResearchIME~\cite{buschek2018researchIME} enabled naturalistic observation of typing patterns, autocorrection use, and suggestion acceptance outside laboratory settings, revealing systematic individual differences in typing strategies. \citet{dhakal2018observations} identified eight distinct typing profiles that differ in finger usage, speed, and accuracy, with performance strongly predicted by the number of fingers employed. Even ambient factors matter: \citet{mecke2025exploring} found that background music tempo influences both typing speed and error rates, highlighting the sensitivity of keystroke signatures to external conditions.

\subsection{Research Gap: Keystroke Dynamics in Human-LLM Collaboration}

Despite this extensive literature, a critical gap remains: no prior work has investigated how keystroke dynamics manifest during LLM-assisted tasks, or whether they can predict collaboration effectiveness. Existing studies have focused on solo text production (essay writing, email composition) or specialized domains (programming, authentication), but human-LLM interaction introduces fundamentally different dynamics. Users engage in iterative prompt refinement, evaluate and adapt to AI outputs, and adjust their input strategies based on response quality. The relationship between typing effort and task success may differ substantially from traditional writing contexts, where output quality is directly produced by the user rather than co-created with an AI system.

Moreover, it remains unclear whether the cognitive effort users invest in prompting will relate to their satisfaction with AI outputs. 
In practice, this relationship may be indirect or even counterintuitive: after unsatisfactory replies, users may, for example, put more effort into subsequent prompts or exhibit changes in affect.
Prior work has documented dissociations between process effort and outcome quality in human-AI collaboration (Section 2.1), but no research has examined whether keystroke-based indicators of user effort align with perceived usefulness of AI responses.
Our work addresses these gaps by systematically examining keystroke behavior during LLM interaction across varying task difficulty and device types, while investigating both the sensitivity of keystroke metrics to cognitive demand and their validity as predictors of collaboration success.

\section{User Study}

Prior research suggests that typing behavior is linked to diverse user signals, such as cognitive effort and affective state, making it a valuable behavioral metric \cite{acien2022detection, trojahn2013emotion}. Thus, keystrokes offer a promising way to continuously assess the success of human-AI collaboration. Based on this motivation, our work is guided by the following research questions:

\begin{itemize}
    \item \textbf{RQ1}: Can we use keystroke dynamics as continuous evaluation metrics of human-AI collaboration across device types?
    \item \textbf{RQ2}: How does typing behavior vary across different task complexities and device types when interacting with LLMs?
    \item \textbf{RQ3}: Can typing behavior indicate the perceived usefulness of LLM outputs?
\end{itemize}

To investigate these questions, we designed an experiment in which participants interacted with an LLM while their keystroke dynamics and subjective evaluations were recorded. The following sections describe the experimental conditions, tasks, and system setup.

\subsection{Study Design}

We conducted a between-subjects experimental design with two independent variables: Device (two levels: Mobile and Desktop) and Task Difficulty (two levels: Easy and Hard). Participants were required to iteratively refine their prompts in an AI-assisted task based on how well the AI-generated response fulfilled the task's predefined requirements.

\subsection{Sample Size Justification}
We conducted an a priori power analysis using G*Power (version 3.1.9.7) to determine the required sample size for our 2 (Setting: Mobile vs. Desktop, between-subjects) $\times$ 2 (Difficulty: Easy vs. Hard, within-subjects) mixed factorial design. We aimed to detect a medium-sized interaction effect between Setting and Difficulty. Based on a meta-analysis of typing experiments in HCI~\cite{obukhova2021meta}, a medium effect size for within-subject designs is estimated as Hedges's $g_\mathrm{rm} = .36$, which corresponds approximately to Cohen's $f \approx .25$, following \citet{yatani2016effect}. Using this effect size, a significance level of $\alpha = .05$, and a desired statistical power of $1 - \beta = .80$, the power analysis indicated a minimum total sample size of 34 participants (17 per group). To ensure proper counterbalancing of within-subject order (i.e., Easy–Hard vs. Hard–Easy), we opted to recruit 36 participants (18 per group), allowing an equal distribution across counterbalanced orders. In the G*Power setup, the number of groups was set to 2 (Mobile vs. Desktop), and the number of measurements to 2 (Easy and Hard). We assumed a correlation among repeated measures of .5 and a nonsphericity correction $\epsilon = 1$, consistent with the assumptions and empirical data from prior work~\cite{obukhova2021meta}. The study was approved by the local ethics review board.

\subsection{Dependent Variables}
To evaluate the effects of device type and task difficulty on user typing behavior and perceived LLM output utility, we collected a range of dependent variables categorized under two main dimensions: Interaction Behavior and User Experience. These variables were selected to capture both objective performance measures and subjective experiences. 

\subsubsection{Interaction Behavior.}
These variables capture how users interact with the system during the task, focusing on the typing behavior. They provide insights into participant engagement, cognitive effort, and the effectiveness of user-AI interactions.
Typing behavior metrics are informative of users' cognitive effort and writing fluency~\cite{conijn2019understanding, brizan2015utilizing, oliveira2020writing}. We included keystroke measures such as keystrokes per prompt, words per prompt, and inter-key interval (IKI), which are often associated with hesitation or cognitive load~\cite{oliveira2020writing}. We additionally included the number of pauses, which are considered significant time gaps between keystrokes (e.g., exceeding 1000ms) and have been shown to indicate user performance~\cite{shrestha2022pausing}.
We furthermore included the frequency of backspace \add{and delete key} usage. 
\add{While there are other common error-related typing features, such as corrected error rate, uncorrected error rate, and total error rate~\cite{arif2009error, soukoreff2003text}, these measures are primarily designed to assess transcription accuracy in controlled typing tasks. Since our study involves open-ended prompt formulation, the amount of backspace/delete key usage was chosen to capture corrections.}

\subsubsection{User Experience}
Participants were asked to rate the following three statements after every LLM response on a scale from strongly disagree to strongly agree: "The AI's response was useful.", "Refining the AI's response was difficult.", "Refining the AI's response was mentally demanding.".
At the end of every task, participants additionally filled out a raw NASA-TLX questionnaire and rated the following statement on a scale from strongly disagree to strongly agree: "I am confident that the final meal plan meets all the requirements.".

\subsection{Task}

We used two levels of task complexity to study how users adapt their typing behavior under different cognitive demands.
The \textbf{easy task} was designed to require some interaction with the LLM, but should be solvable with minimal effort, \add{e.g., 2-4 prompts}.
The \textbf{hard task} included more complex and layered requirements, \delete{increasing cognitive effort. We expected that the hard task would lead to multiple prompt refinements, potentially causing participant frustration with the LLM performance.}\add{designed to lead the LLM to produce output that does not fulfill the hard task's criteria on the first few prompts, thus creating the desired situation of multiple back-and-forths (e.g., 5-10 prompts) where the user must allocate more mental effort in thinking about the prompt formulation based on how past prompts evoked past responses and which task requirements are already fulfilled or not. We expected the hard task to increase user mental effort, reflecting the difficulty of solving complex tasks through human-LLM interaction, in our study, artificially induced by exploiting known model limitations.}

To select a suitable task for our experiment and task complexity levels, we first defined a set of requirements. The task needed to encourage prompt iteration through multiple rounds of input refinement, allow difficulty manipulation through adjustable constraints, include clear and verifiable success criteria, be feasible on both desktop and mobile devices, and remain accessible to a broad audience without specialized knowledge.
We finally selected the task of \textbf{generating a 7-day meal plan with dietary constraints of varying complexity}. Participants interacted with a large language model to create the plan and revised their prompt based on the model's output. Task difficulty was manipulated by adjusting the number of dietary requirements. 

To define the dietary constraints for the task, we investigated the common limitations of large language models. Several recent studies~\cite{fu2024largelanguagemodelsllms, xu2025llmgeniusparadoxlinguistic, qiu2023largelanguagemodelstemporally} highlight areas where LLMs typically struggle, including \add{but not limited to} counting and maintaining item quantities, avoiding repetitions, and performing temporal planning. \add{Prior work suggests that LLMs' deficiency in solving counting tasks stems from their design, namely their probabilistic nature, sequential token prediction, and byte-level tokenization~\cite{zhang2024counting}.} These models also face challenges in respecting constraints that require internal logic, correcting their own mistakes, and providing accurate numerical distributions, such as percentages that correctly sum to 100\% \add{or keeping a precise count}. 

We designed dietary requirements (see \autoref{tab:requirements}) that map directly to these LLM limitations to make the hard task more challenging, causing high user effort and potential distress. Requirements were split into an easy and a hard condition to study how users adapt their typing behavior under different cognitive demands. The choice of LLM also influences how difficulty is experienced in practice, as different models vary in their capabilities. The design of the easy and hard tasks was refined through pilot testing with the LLM model and real users to achieve the desired difficulty levels. The final task instructions used in the study can be found in \autoref{appendix:task-instructions}. 

\begin{table}[h]
\centering
\footnotesize
\caption{Dietary requirements for the easy and hard task conditions.}
\label{tab:requirements}
\begin{tabular}{p{10cm}cc}
\toprule
\textbf{Dietary requirements} & \textbf{Easy task} & \textbf{Hard task} \\
\midrule
7-day meal plan & \checkmark & \checkmark \\
5 days: 3 meals + 1 snack per day & \checkmark & \checkmark \\
2 days: 2 meals only & \checkmark & \checkmark \\
Dairy-free & \checkmark & \checkmark \\
No processed food & \checkmark & \\
No repeated dishes &  & \checkmark \\
No repetition of main ingredients on 2 consecutive days &  & \checkmark \\
Even distribution of calories: ~2000 kcal/day &  & \checkmark \\
High in protein, low in carb meal plan &  & \checkmark \\
Daily macronutrient breakdown (\% carbs, fats, proteins must add up to 100\%) &  & \checkmark \\
\bottomrule
\end{tabular}
\end{table}

\subsection{Apparatus}
To conduct the user study, we developed a local web-based platform for controlled human-LLM interaction. The system operated entirely offline, comprising a React frontend, a Python FastAPI backend, and a local SQLite database. This client-server architecture ensured low latency and complete data privacy. The frontend, designed to be familiar to users of common AI chatbots, is fully responsive for both desktop and mobile devices. To maintain a consistent experimental environment, we implemented several safeguards, such as disabling copy-paste, autocorrect, and browser spellcheck on all text inputs and including warnings to prevent accidental page reloads.

The platform integrated a local instance of the Llama 3.2 (3B) language model hosted via the Ollama server. 
We intentionally selected a smaller model to ensure the experimental tasks required iterative prompt refinement from participants and to emulate the constraints of privacy-preserving on-device LLMs used in mobile settings.
All communication with the LLM occurred directly between the frontend and the local server to minimize latency. A system prompt was injected at the start of each session to standardize model outputs to metric units, and a chat history was maintained to provide conversational context. Responses were streamed to the user interface to simulate a real-time interaction.

A custom keystroke logging mechanism was implemented in the frontend to capture detailed typing behavior. It recorded every keypress event during prompt entry, logging the key, a high-precision timestamp, cursor position, and the full input text at that moment. All experimental data, including participant demographics, task evaluations (NASA-TLX), interaction logs, and fine-grained keystroke data, were sent to the backend and stored in the SQLite database. The relational database schema linked these data points, enabling a comprehensive analysis of user behavior in relation to task conditions and subjective feedback.

\subsection{Procedure}

\begin{figure}[ht!]
  \centering
  \includegraphics[width=\textwidth]{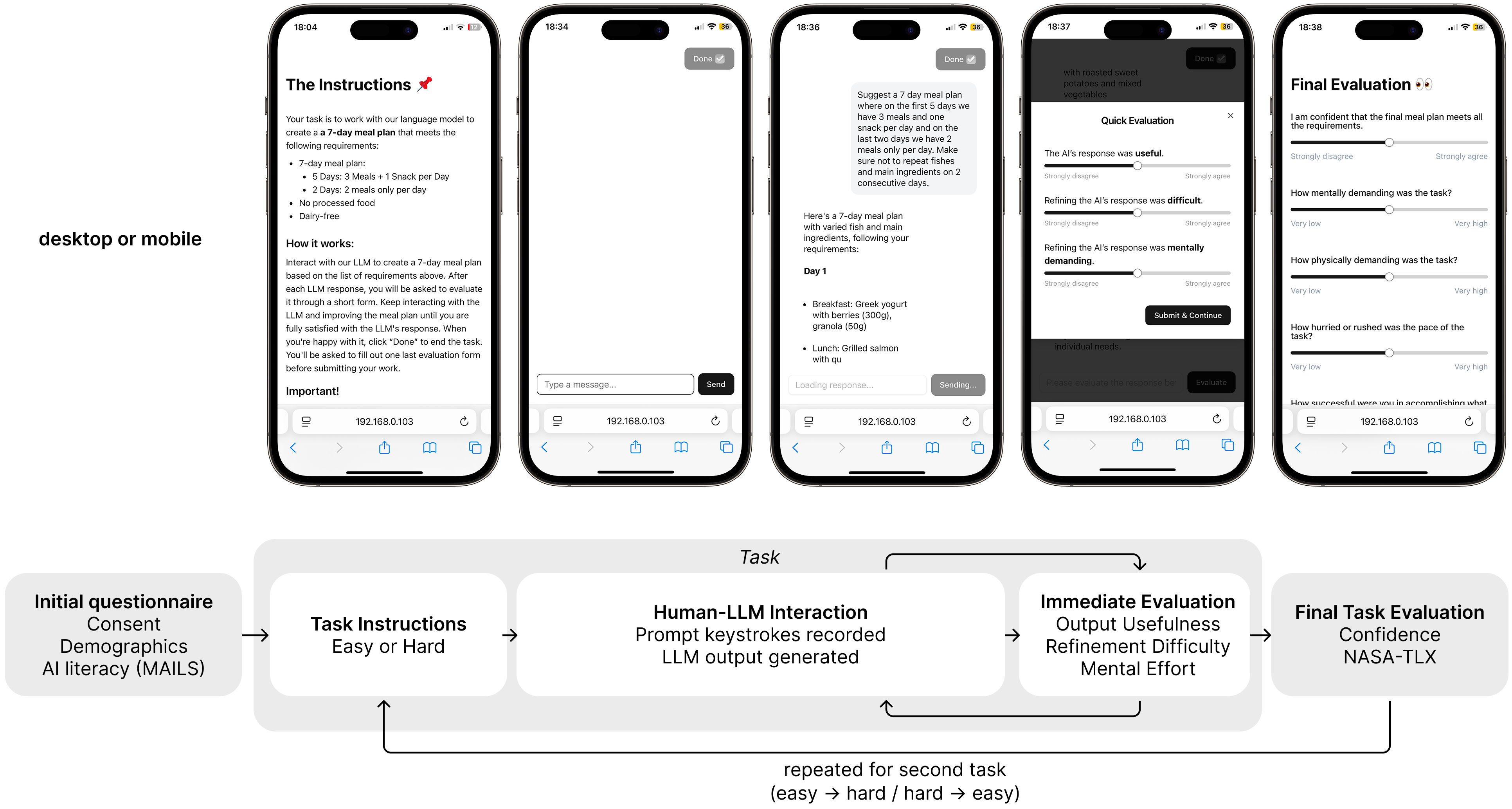}
  \caption{Experiment protocol of the between-subjects user study. Participants either completed the desktop or mobile condition. Every participant completed an easy and a hard task in counterbalanced order. The mobile UI is shown to exemplify the interaction.}
  \Description{A diagram illustrating the study interface and protocol. The top section displays five sequential mobile phone screenshots showing the user flow: 1) Task instructions for creating a 7-day meal plan, 2) The chat interface input field, 3) The LLM response generation, 4) A 'Quick Evaluation' pop-up form asking about usefulness and mental demand, and 5) A 'Final Evaluation' form with slider scales. The bottom section shows a flowchart of the procedure, starting with an 'Initial questionnaire' (Consent, Demographics, AI literacy). This leads to the main 'Task' block, which contains a loop: 'Task Instructions' leads to 'Human-LLM Interaction' (recording keystrokes), followed by 'Immediate Evaluation.' The evaluation loops back to interaction or proceeds to 'Final Task Evaluation' (Confidence, NASA-TLX). The diagram notes that the task block is repeated for a second task, counterbalancing easy and hard difficulties.}
  \label{fig:study_protocol}
\end{figure}

A figure outlining the experiment flow and web-based interface can be found in \autoref{fig:study_protocol}.
Participants completed the study in a quiet room, where they first received a brief introduction to the research goals and provided informed consent. Each person was assigned to either a desktop (27-inch monitor, Apple Magic Keyboard) or mobile (iPhone 15 Pro) condition. Both the device type and the task difficulty order (easy → hard or hard → easy) were counterbalanced. The web-based experiment began with initial questionnaires covering demographics and AI literacy via the MAILS scale \cite{carolus2023mails}. This was followed by a baseline typing task, where participants wrote a simple prompt in response to a neutral scenario to capture their natural typing behavior.

The main procedure was repeated for both an easy and a hard task. For each task, participants received printed instructions and could ask for clarification before starting. \add{All participants received the exact same task description to ensure a consistent starting baseline.} They then engaged in a chat with the LLM to complete the assignment. After each AI-generated response, they were required to rate its usefulness, as well as the difficulty and mental effort they experienced while writing the prompt. After every interaction, participants could decide to either continue the conversation or conclude the task. Upon finishing a task, they completed a NASA-TLX questionnaire to assess workload and answered a question about their confidence. The final evaluation after the second task included additional questions on keyboard familiarity and preferred typing language.
All participants received a fixed compensation of 12 euros/hour for their participation.

\subsection{Participants}

Our study involved 36 participants (58.3\%  identified as male and 41.7\% as female) with an age range of 19 to 34 (M=24.5). Most participants held a Bachelor's degree (52.8\%), followed by a high school diploma (30.6\%) and a Master's degree (13.9\%).
Participants were experienced and frequent users of large language models (LLMs). A significant majority (69.4\%) reported using LLMs multiple times a day. Their primary use cases were for learning, research, coding, and writing. When asked about their preferred device, participants showed a strong inclination towards desktop usage. On a scale from 0 (fully mobile use) to 20 (fully desktop use), the average score was 14.4 (SD=4.8).
To gauge domain knowledge for our diet-planning task, we asked participants to rate their nutrition expertise on a 0 (none) to 20 (expert) scale. The average self-assessment was moderate, with a mean score of 10.2 (SD=4.9).
Finally, we assessed AI literacy using the short version of the Meta AI Literacy Scale (MAILS) \cite{carolus2023mails}. The results indicated that participants felt most confident in using AI to solve real-world problems. Conversely, they reported lower confidence in more technical areas, such as programming or designing AI systems.

\subsection{Analysis}
\label{sec:analysis}
Once the user study was completed, we proceeded with the data analysis phase. Before conducting any analysis, we performed data cleaning and preprocessing. The data from two participants were discarded: one due to technical issues, and the other due to insufficient English proficiency.
From the raw keystroke logs, we computed several behavioral metrics, including typing speed, input length, pause count, and backspace usage (see \autoref{tab:typing_metrics}). We computed these metrics at all levels: per participant, per condition (easy/hard, desktop/mobile), and per interaction. This allowed us to compare typing patterns across different groups.
All plots and analysis scripts were written in Python using libraries such as \texttt{pandas}, \texttt{seaborn}, and \texttt{matplotlib}. All data and analysis scripts are available at \href{https://osf.io/dsnqf}{https://osf.io/dsnqf}. 

\begin{table}[t]
\centering
\footnotesize
\caption{\add{Analyzed keystroke dynamics and descriptions}}
\label{tab:typing_metrics}
\begin{tabular}{l p{9 cm}}
\toprule
\add{\textbf{Metric}} & \add{\textbf{Description}} \\ 
\midrule
\add{Keystroke count} & \add{Number of keystrokes per prompt} \\ 
\add{Words per prompt (WPP)} & \add{Number of words per prompt} \\ 
\add{Pause count} & \add{Number of pauses (>1000 ms inter-key intervals) per prompt } \\ 
\add{Inter-key interval (IKI)} & \add{Time difference (ms) between consecutive key-down events} \\ 
\add{Backspace usage} & \add{Number of backspace/delete key events normalized by total keystroke count} \\
\bottomrule
\end{tabular}
\end{table}

To investigate the effects of task difficulty and device type on typing and user experience measures, we fitted linear mixed-effects models (LMMs) using the \texttt{lme4} package. 
\add{LMMs allow us to analyze hierarchical data by accounting for both fixed effects (e.g., condition) and random effects (e.g., individual user baseline differences). This provides a more robust analysis than standard summary statistics, which assume data independence~\cite{bates2015fitting}. Prior to model fitting, we verified LMM assumptions separately for each 
dependent variable. Normality of residuals was assessed via visual inspection 
of Q-Q plots, and homoscedasticity was evaluated through residual-versus-fitted 
and scale-location plots. Mental Demand residuals were approximately normally 
distributed. For the four keystroke count metrics (Keystroke Count, WPP, Pause 
Count, and IKI), residuals exhibited mild positive skew, consistent with the 
non-negative, count-like nature of these variables and the presence of 
occasional outlier interactions. No systematic patterns indicating 
heteroscedasticity were identified across conditions. Given the robustness of 
LMMs to moderate violations of normality, particularly in larger 
datasets~\cite{bates2015fitting}, we proceeded with untransformed variables 
to preserve interpretability of the coefficients. For the IKI model, a boundary 
singular fit was detected (condition\_order variance $= 0$), and is reported 
transparently in the results.
}
Models included random intercepts for user ID and condition order. 
Model comparisons were based on the Akaike Information Criterion (AIC), a standard measure for model quality that balances fit and complexity.
To explore whether typing behavior predicts perceived usefulness of AI feedback, we employed three modeling approaches: linear mixed models, principal component regression (PCR), and random forest regression. Prior to modeling, predictors were standardized to improve interpretability. All analyses were implemented in R using libraries such as \texttt{dplyr}, \texttt{ggplot2}, \texttt{lmerTest}, and \texttt{report}.

\section{Results}

In the following, we report the interaction behavior and user experience results from the user study. Results on the prediction of perceived usefulness of AI output from keystroke metrics are reported as well.

\subsection{Interaction Behavior}

\subsubsection{Number of Interactions}
\begin{figure}[H]
  \centering
  \includegraphics[width=0.7\textwidth]{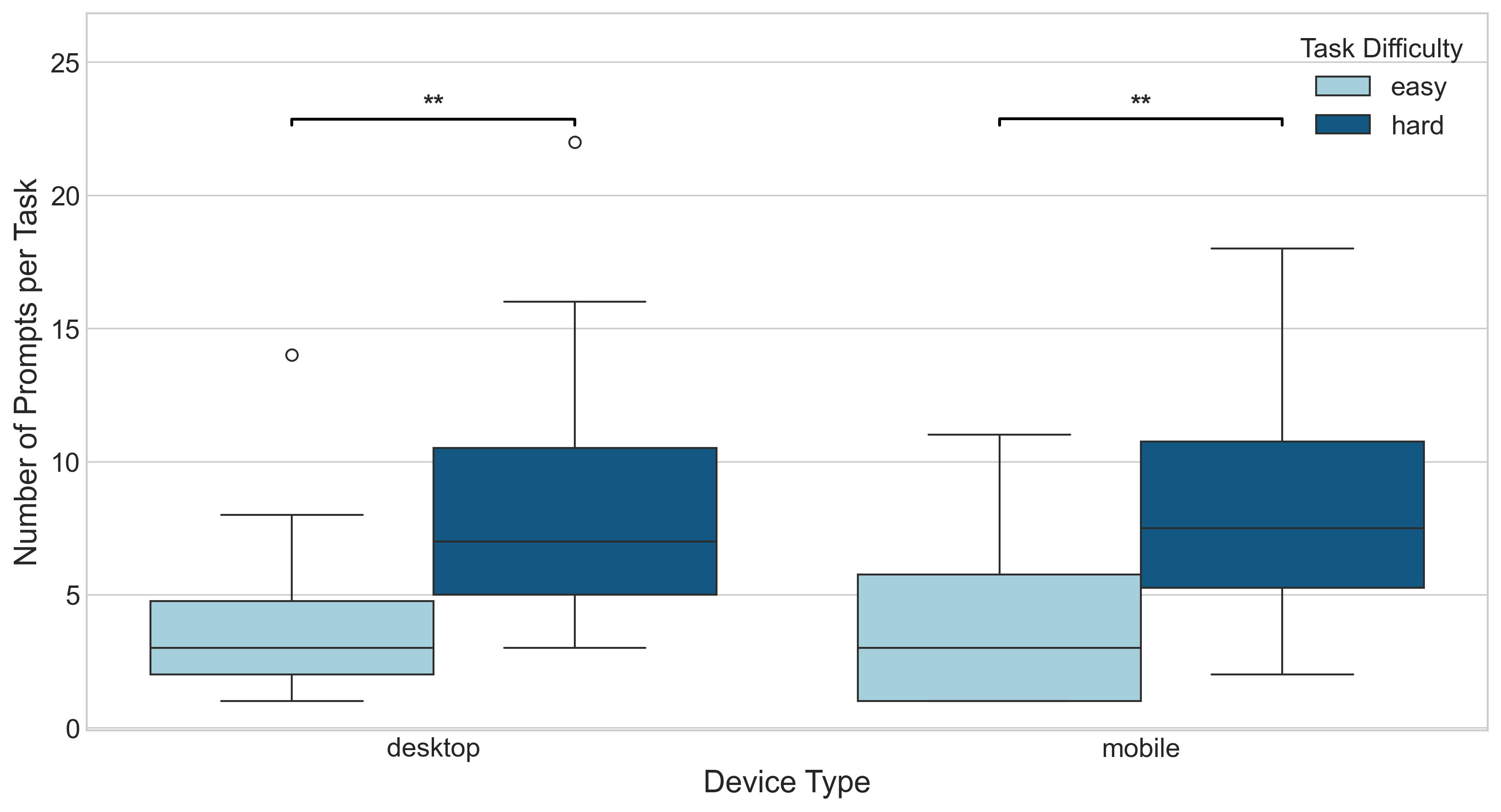}
  \caption{Interaction counts (number of prompts) by device type and task.}
  \Description{This box plot illustrates the number of prompts per task across desktop and mobile devices, categorized by easy (light blue) and hard (dark blue) task difficulty. The data reveals that task difficulty is a primary driver of interaction volume; for both device types, hard tasks require a significantly higher median number of prompts compared to easy tasks. Statistical significance brackets with double asterisks confirm that these differences are significant across both devices.}
  \label{fig:interactions-per-condition}
\end{figure}

On average, participants spent approximately 46.6 minutes interacting with the experimental system. Each participant completed two tasks, categorized as \textit{easy} and \textit{hard}. 
Across all participants, a total of 102,454 keystrokes were recorded, with a mean of 2,845.9 keystrokes per participant. Additionally, 436 human-AI interactions were captured, corresponding to an average of 12.1 interactions per participant.
When broken down by task type, \textit{easy} tasks averaged 3.8 interactions per participant, while \textit{hard} tasks averaged 8.3 interactions.
A visual breakdown is presented in \autoref{fig:interactions-per-condition}.
There was a significant main effect of task difficulty on interaction count (p < 0.001). Hard tasks resulted in significantly more interactions than easy tasks (p < 0.01), regardless of device type. There was no significant main effect of device type on the number of interactions, nor a significant interaction effect.

\subsubsection{Keystroke Metrics}

\begin{figure*}[h]
  \centering
  
  \begin{subfigure}{0.48\linewidth}
    \centering
    \includegraphics[width=\linewidth]{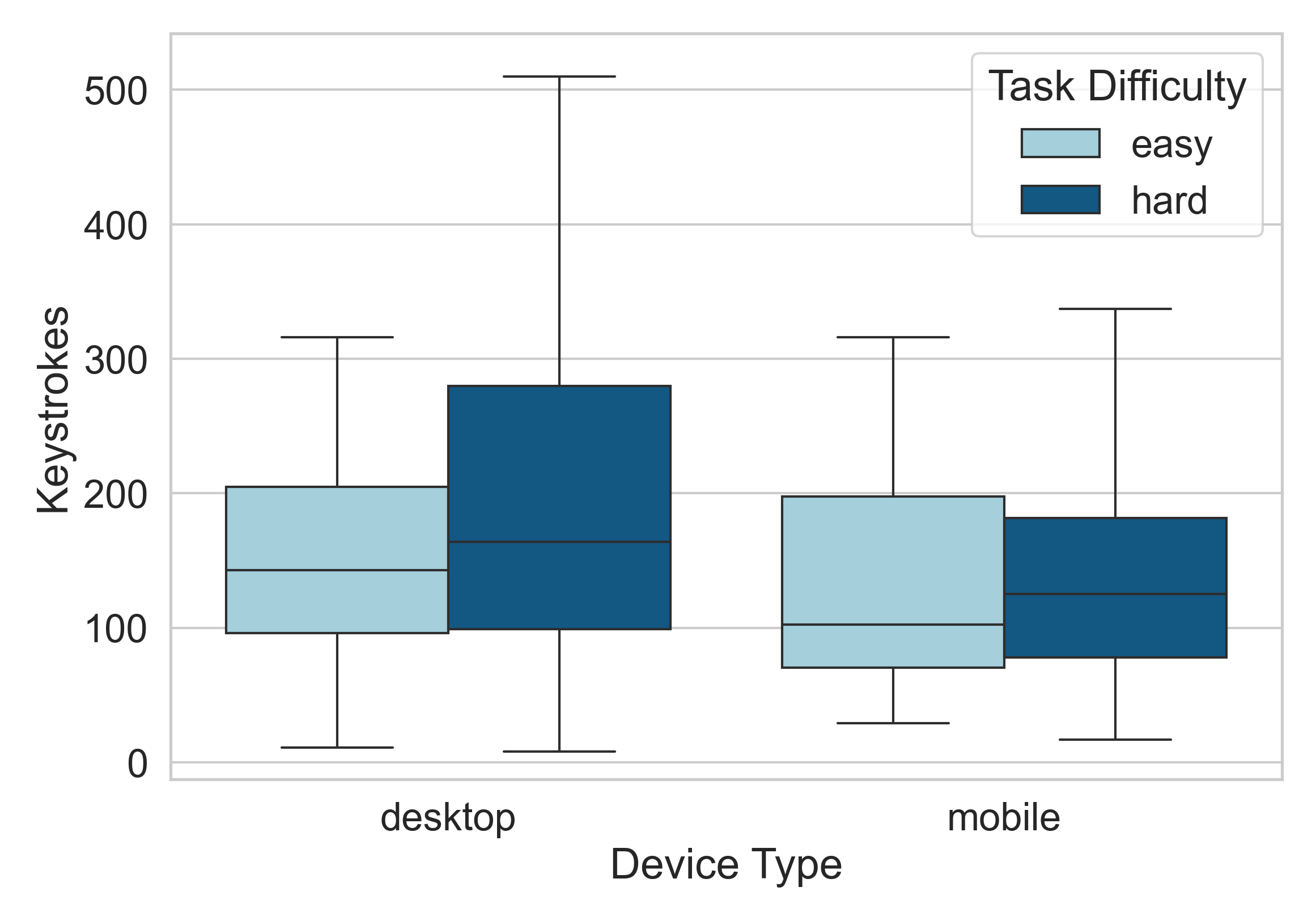}
    \label{fig:top_left}
  \end{subfigure}
  \hfill 
  \begin{subfigure}{0.48\linewidth}
    \centering
    \includegraphics[width=\linewidth]{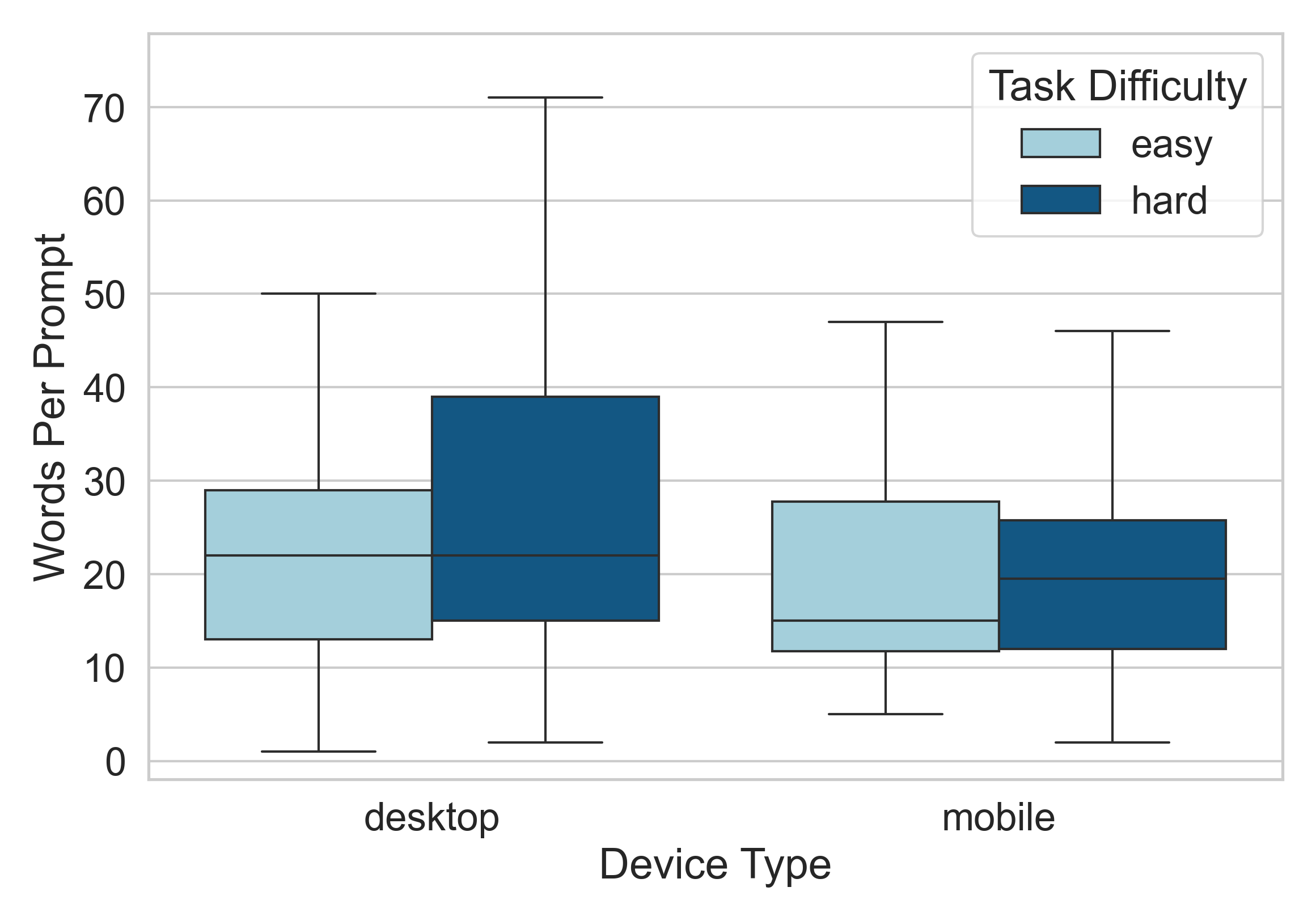}
    \label{fig:top_right}
  \end{subfigure}
  \vspace{-1em}
  \begin{subfigure}{0.48\linewidth}
    \centering
    \includegraphics[width=\linewidth]{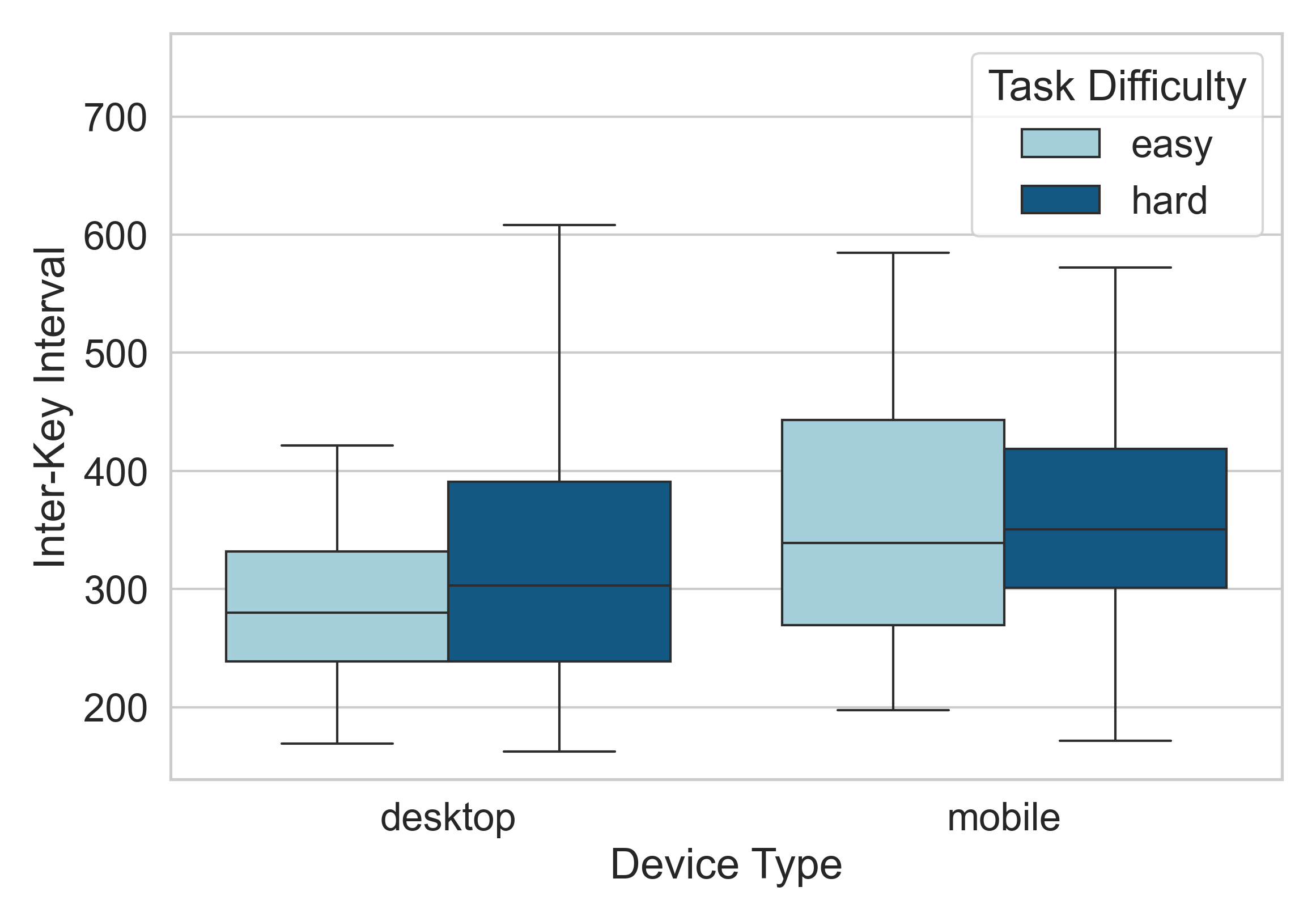}
    \label{fig:bottom_left}
  \end{subfigure}
  \hfill 
  \begin{subfigure}{0.48\linewidth}
    \centering
    \includegraphics[width=\linewidth]{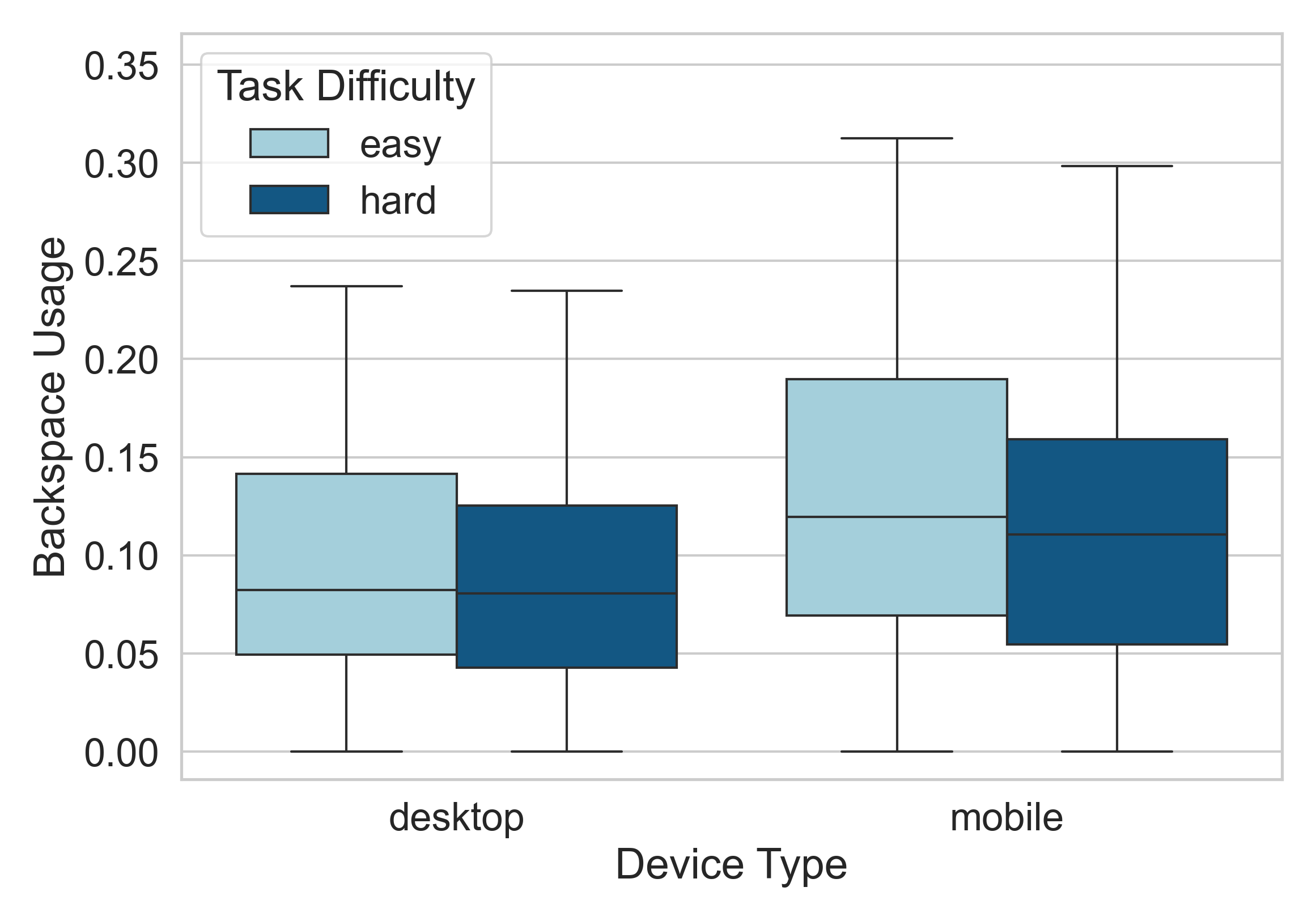}
    \label{fig:bottom_right}
  \end{subfigure}
  
  \caption{Box plots of keystroke metrics by task difficulty and device type.}
  \Description{This figure presents a grid of four box plots comparing keystrokes, words per prompt, inter-key interval, and backspace usage across desktop and mobile devices for easy and hard tasks. Mobile interactions show significantly slower typing speeds (higher inter-key intervals) and more frequent error correction (higher backspace usage).}
  \label{fig:keystroke_plots}
\end{figure*}

\add{To investigate how task difficulty and device type influenced typing behavior, 
we analyzed five key metrics: Keystroke Count, Words Per Prompt (WPP), Pause 
Count, Inter-Key Interval (IKI), and Backspace Usage. For each metric, we 
compared three LMM specifications of increasing complexity via AIC: a base 
model with random intercepts for participant only 
(\textit{model\textsubscript{1}}: \texttt{Outcome $\sim$ task\_difficulty 
$\times$ device\_type + (1 | user\_id)}), a model additionally including 
condition order (\textit{model\textsubscript{2}}: \texttt{+ (1 | 
condition\_order)}), and a full model further adding prompt order 
(\textit{model\textsubscript{3}}: \texttt{+ (1 | prompt\_order)}). Fixed 
effects were identical across all three specifications: task difficulty, device 
type, and their interaction.
}
~\autoref{tab:typing-metrics-summary} reports, for each metric, the selected random effects structure, AIC of the best-fitting model, marginal $R^2$ (variance explained by fixed effects alone), and conditional $R^2$ (variance explained by the full model including random effects) 
\cite{nakagawa2013general}, alongside the key fixed-effect coefficients. 
~\autoref{fig:keystroke_plots} visualizes the distributions of these metrics by condition.

\add{
For four of five metrics (Keystroke Count, WPP, Pause Count, IKI), 
\textit{model\textsubscript{3}} was selected as best-fitting by AIC, indicating 
that both condition order and prompt order account for meaningful variance 
beyond individual differences. For Backspace Usage, AIC favored the simpler 
\textit{model\textsubscript{1}} structure ($\Delta$AIC $=$ 2.00 vs.\ 
\textit{model\textsubscript{2}}), suggesting that correction behavior does not 
vary systematically with prompt or condition order. Prior to examining individual fixed effects, we note that LMM assumptions  were verified for each metric as described in Section~\ref{sec:analysis}.  Residuals showed mild positive skew for count-based metrics (Keystroke Count,  WPP, Pause Count, IKI), consistent with the distributional properties of  typing data~\cite{dhakal2018observations}, but no violations of homoscedasticity were detected. Results should be interpreted accordingly.
}

\add{
Across all metrics, marginal $R^2$ values were modest (range: .02\,--\,.05), 
consistent with the high degree of individual variability in typing behavior 
documented in prior work~\cite{dhakal2018observations, yang2025identify}. 
Conditional $R^2$ values were substantially higher (.35\,--\,.48), confirming 
that individual differences captured by the random effects structure account 
for meaningful variance and justify the mixed-effects approach over simpler 
alternatives. One exception: the IKI model exhibited a boundary singular fit, 
with the condition\_order random effect variance collapsing to zero. 
$R^2_{\text{conditional}}$ is therefore not reported for this metric, and 
interpretation is restricted to fixed effects.
}

\begin{table*}[t]
\footnotesize
\centering
\caption{Summary of linear mixed model results for keystroke behavior metrics. \add{Random effects: U=user, O=order, P=prompt; AIC = goodness of fit of best-fitting model; $R^2_{\text{marg}}$ = variance explained by fixed effects; $R^2_{\text{cond}}$ = variance explained by fixed and random effects combined~\protect\cite{nakagawa2013general}; $\hat{\beta}$ = fixed effects for task difficulty, device type, and task difficulty $\times$ device type.}}
\label{tab:typing-metrics-summary}
\begin{tabular}{llrrrrrr}
\toprule
\textbf{\scriptsize{Metric}}
  & \textbf{\scriptsize{\add{Random Effects}}} 
  & \textbf{\scriptsize{\add{AIC}}} 
  & $\boldsymbol{\scriptsize{\add{R^2_{\text{marg}}}}}$ 
  & $\boldsymbol{\scriptsize{\add{R^2_{\text{cond}}}}}$
  & \textbf{\scriptsize{Task (hard) $\hat{\beta}$}} 
  & \textbf{\scriptsize{Device (mobile) $\hat{\beta}$}} 
  & \textbf{\scriptsize{Task $\times$ Device $\hat{\beta}$}} \\
\midrule
Keystrokes
  & U+0+P 
  & 5841.08 & .05 & .44 
  & $127.89^{***}$ 
  & $-5.04$~(n.s.) 
  & $-88.41^{*}$ \\

WPP             
  & U+0+P 
  & 4122.87 & .05 & .41
  & $16.38^{***}$ 
  & $-2.74$~(n.s.) 
  & $-10.34$~(n.s.) \\

Pauses    
  & U+0+P 
  & 3455.56 & .02 & .48 
  & $6.27^{***}$ 
  & $0.23$~(n.s.) 
  & $-3.06$~(n.s.) \\

IKI             
  & U+0+P$^{\dagger}$ 
  & 5506.07 & .05 & ---
  & $46.93^{*}$ 
  & $83.93^{*}$ 
  & $-58.55^{*}$ \\

Backspaces 
  & U 
  & $-$1146.65 & .04 & .35
  & $-0.005$~(n.s.) 
  & $0.031$~(n.s.) 
  & $-0.002$~(n.s.) \\
\bottomrule
\multicolumn{8}{l}{%
  Significant effects: $^{*}p < .05$;\; $^{**}p < .01$;\; $^{***}p < .001$.} \\
\end{tabular}
\end{table*}

\paragraph{Keystrokes Count.} 
Participants produced significantly more keystrokes on hard tasks, with an average increase of 127.89 keystrokes per interaction compared to easy tasks ($SE = 29.27$, $p < .001$, 95\% CI $[70.35, 185.42]$). Device type had no main effect on total keystroke count ($\beta = -5.04$, $SE = 47.83$, $p = .916$), indicating that typing output volume did not differ between mobile and desktop users in isolation. However, a significant interaction effect revealed that the increase in keystrokes under hard conditions was notably attenuated on mobile devices ($\beta = -88.41$, $SE = 41.65$, $p = .034$), suggesting that input constraints may have limited the amount of text generated during high-difficulty tasks on mobile.

\paragraph{Words Per Prompt (WPP)}  
Task difficulty significantly influenced prompt verbosity, with hard tasks leading to a mean increase of 16.38 words compared to easy tasks ($SE = 4.00$, $p < .001$, 95\% CI $[8.51, 24.25]$). The main effect of device type was non-significant ($\beta = -2.74$, $SE = 5.64$, $p = .627$), suggesting that participants produced similar prompt lengths across desktop and mobile platforms. Similarly, the interaction between task difficulty and device type was not significant ($\beta = -10.34$, $SE = 5.70$, $p = .070$), although the negative direction may indicate a potential attenuation of verbosity under high difficulty on mobile. Overall, task difficulty was the dominant predictor of prompt length.

\paragraph{Pause Count.} 
The number of pauses per interaction was significantly higher during hard tasks, with an estimated increase of 6.27 pauses compared to easy tasks ($SE = 1.84$, $p < .001$, 95\% CI $[2.65, 9.88]$). This suggests that participants engaged in more reflective or effortful typing when cognitive demand increased. There was no significant main effect of device type ($\beta = 0.23$, $SE = 2.73$, $p = .934$), nor a significant interaction with task difficulty ($\beta = -3.06$, $SE = 2.62$, $p = .243$), indicating that pause behavior was not notably influenced by device platform.

\paragraph{Inter-Key Interval (IKI)} 
The inter-key interval (IKI), a proxy for typing fluency, was significantly affected by both task difficulty and device type. Hard tasks led to slower typing, with an average increase of 46.93 ms between keystrokes compared to easy tasks ($SE = 20.21$, $p = .021$, 95\% CI $[7.20, 86.66]$). Typing was also slower on mobile devices ($\beta = 83.93$, $SE = 36.51$, $p = .022$), likely reflecting physical constraints of mobile input. Importantly, there was a significant interaction effect ($\beta = -58.55$, $SE = 28.61$, $p = .041$), indicating that the IKI increase under high task difficulty was less pronounced on mobile, potentially due to a performance ceiling on mobile typing.

\paragraph{Backspace Usage.}  
A linear mixed-effects model revealed no significant main or interaction effects for backspace behavior. Task difficulty did not reliably affect the frequency of backspace usage ($\beta = -0.005$, $SE = 0.009$, $p = .605$), nor did device type show a statistically significant effect ($\beta = 0.031$, $SE = 0.017$, $p = .067$). The interaction between task difficulty and device type was also non-significant ($\beta = -0.002$, $SE = 0.013$, $p = .857$). While there was a trend toward increased editing on mobile, this did not reach significance, and overall, backspace usage remained stable across conditions.

\subsection{User Experience}
\subsubsection{NASA-TLX}
The NASA-TLX questionnaire was used to gather subjective workload assessments across six dimensions: Mental Demand, Physical Demand, Temporal Demand, Performance, Effort, and Frustration. Participants provided ratings on a 21-point scale from 0 to 20, following the standard NASA-TLX procedure. ~\autoref{tab:tlx-summary} summarizes the mean and standard deviation across four experimental conditions: Desktop–Easy, Desktop–Hard, Mobile–Easy, and Mobile–Hard. 
The Raw NASA-TLX, computed as the unweighted average of all six workload subscales further confirms these trends: both hard conditions yielded higher scores (Desktop–Hard: $M = 10.78$, $SD = 3.35$; Mobile–Hard: $M = 10.71$, $SD = 4.11$) compared to their easy counterparts (Desktop–Easy: $M = 6.87$, $SD = 3.84$; Mobile–Easy: $M = 8.01$, $SD = 3.73$).

\renewcommand{\thefootnote}{\fnsymbol{footnote}}
\footnotetext[1]{%
$M$ = mean; $SD$ = standard deviation.
}

\begin{figure*}[ht]
  \centering
  \begin{subfigure}{0.49\linewidth}
    \centering
    \includegraphics[width=\linewidth]{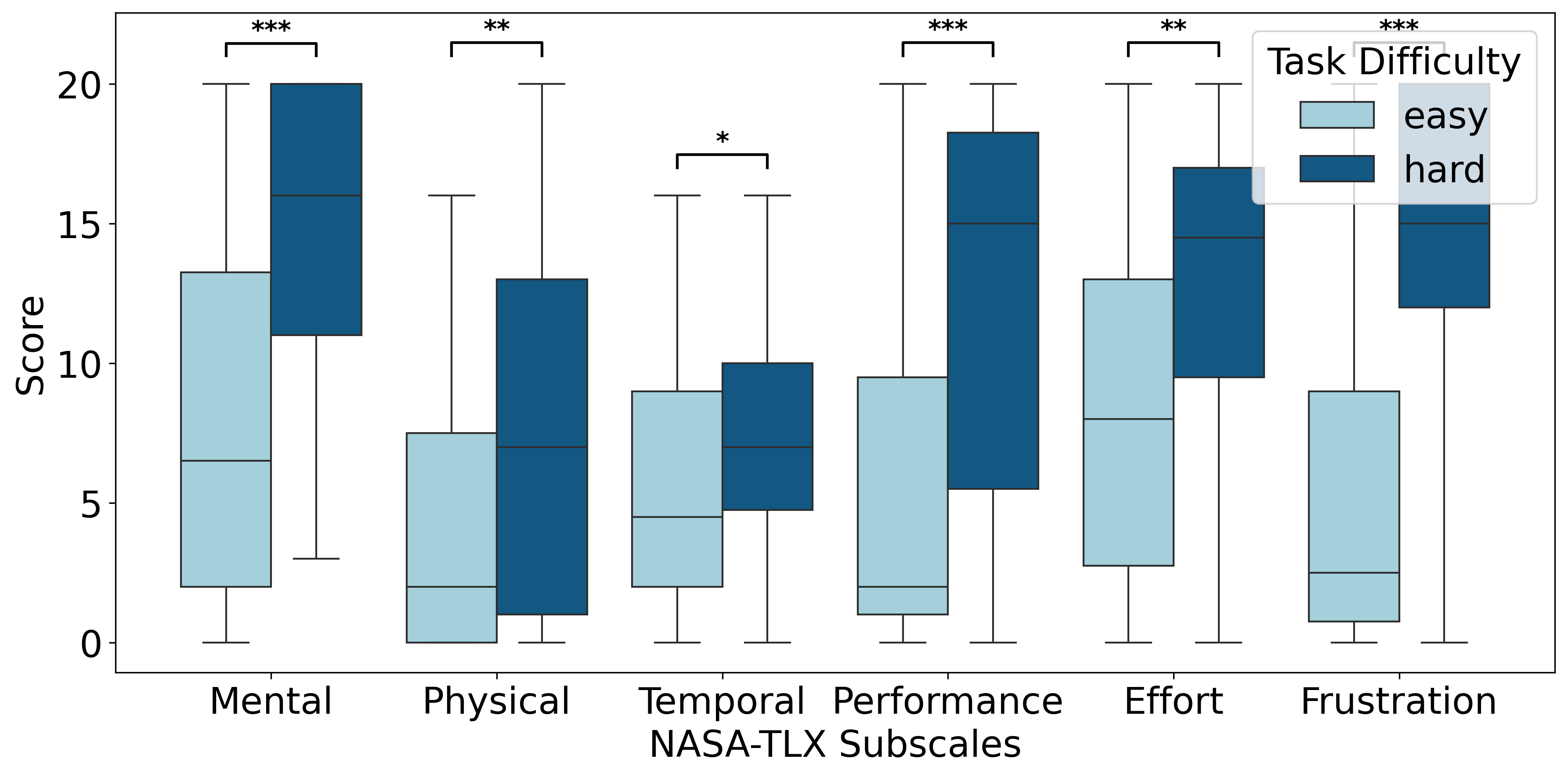}
    \caption{NASA-TLX results by task difficulty}
    \label{fig:tlx_task}
  \end{subfigure}
  \hfill
  \begin{subfigure}{0.49\linewidth}
    \centering
    \includegraphics[width=\linewidth]{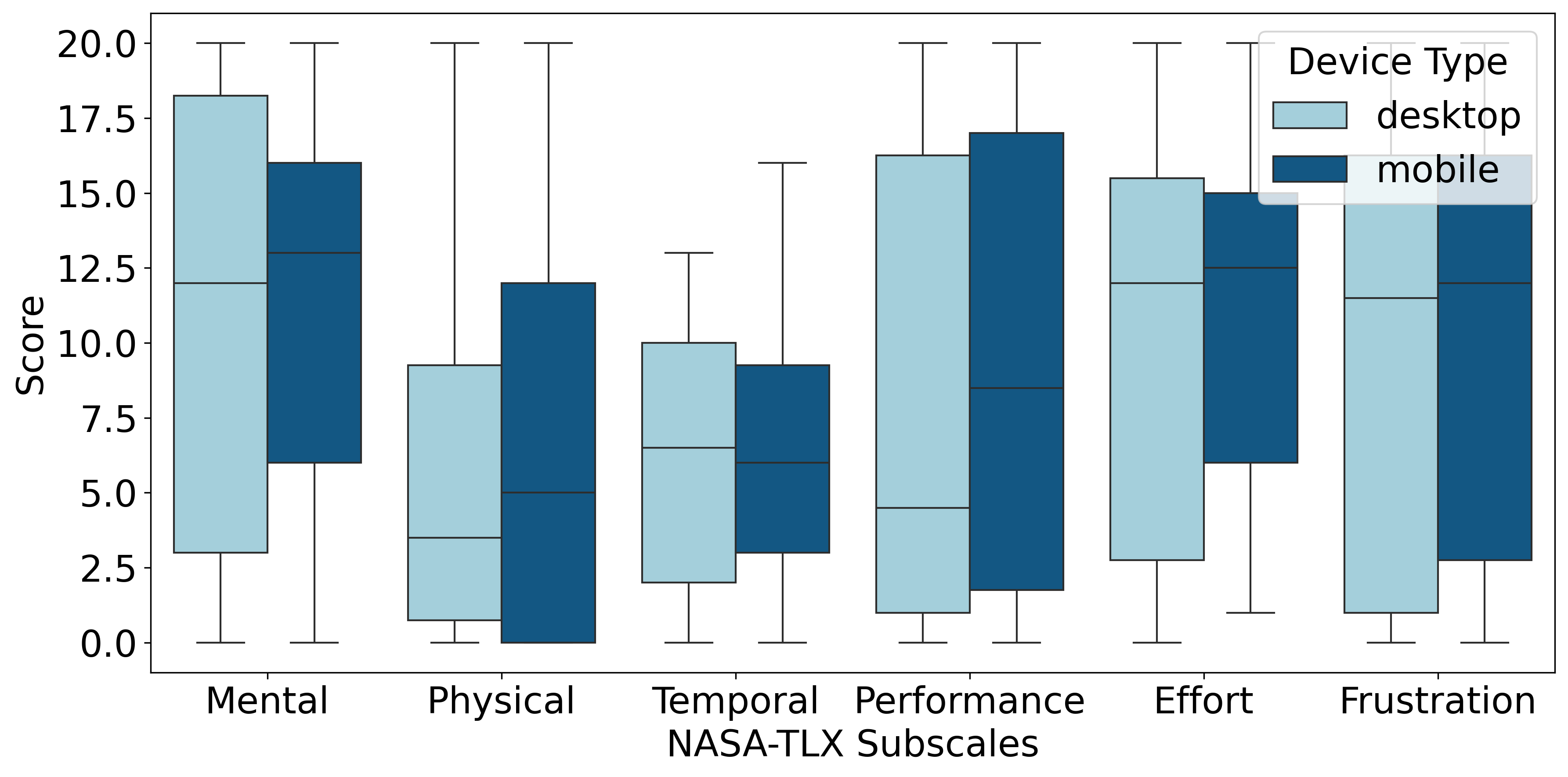}
    \caption{NASA-TLX results by device type}
    \label{fig:tlx_device}
  \end{subfigure}
  \caption{Raw NASA-TLX results by task difficulty and device type for all six subscales: Mental Demand, Physical Demand, Temporal Demand, Performance, Effort, Frustration. Significant differences are marked as follows: p < 0.05 = *, p < 0.01 = **, p < 0.001 = ***.}
  \Description{Figure 5 presents two box plots showing scores for the six NASA-TLX subscales (Mental, Physical, Temporal, Performance, Effort, and Frustration) on a scale of 0 to 20. Plot (a) compares task difficulty, showing that hard tasks (dark blue) consistently result in higher median scores across all subscales compared to easy tasks (light blue), with all differences marked by statistical significance asterisks ranging from p < 0.05 to p < 0.001. Plot (b) compares device type, showing the distribution of scores for desktop (light blue) and mobile (dark blue); in plot (b), the distributions for most subscales overlap, and there are no statistical significance markers indicated.}
  \label{fig:tlx_plots}
\end{figure*}

\begin{table}
\centering
\caption[Raw TLX ratings by condition.]{Raw NASA-TLX ratings of all six subscales and overall by condition (M ± SD).}
\label{tab:tlx-summary}
\footnotesize
\begin{tabular}{lccccc}
\toprule
\textbf{NASA-TLX variable} & \textbf{Desktop – Easy} & \textbf{Desktop – Hard} & \textbf{Mobile – Easy} & \textbf{Mobile – Hard} \\
\midrule
Mental Demand        & 7.00 ± 7.28 & 14.39 ± 6.01 & 8.22 ± 4.86 & 14.39 ± 5.55 \\
Physical Demand      & 2.94 ± 4.30 & 7.44 ± 5.82  & 5.33 ± 5.17 & 8.33 ± 7.44 \\
Temporal Demand       & 4.11 ± 4.30 & 7.94 ± 3.24  & 6.28 ± 4.13 & 7.61 ± 4.82 \\
Performance    & 5.22 ± 7.57 & 11.28 ± 7.19 & 6.72 ± 7.47 & 12.78 ± 7.26 \\
Effort       & 7.50 ± 6.26 & 12.72 ± 6.71 & 8.61 ± 5.40 & 12.78 ± 5.37 \\
Frustration         & 4.89 ± 6.80 & 13.44 ± 6.53 & 6.33 ± 6.02 & 13.94 ± 6.39 \\
\midrule 
Overall raw NASA-TLX         & 6.87 ± 3.84 & 10.78 ± 3.35 & 8.01 ± 3.73 & 10.71 ± 4.11 \\
\bottomrule
\end{tabular}
\end{table}

To further investigate differences in perceived cognitive load, we analyzed \textit{NASA-TLX Mental Demand} ratings using a linear mixed-effects model. The best-fitting model, determined by AIC comparison (AIC = 2147.55), included random intercepts for user ID, condition order, and prompt order. The model explained a substantial amount of variance overall ($R^2_{\text{conditional}} = 0.67$), though the contribution of fixed effects alone was modest ($R^2_{\text{marginal}} = 0.07$).
Results showed a significant main effect of task difficulty: hard tasks were rated as more mentally demanding than easy ones ($\beta = 3.18$, $SE = 0.77$, $p < .001$, 95\% CI $[1.66, 4.70]$). The effect of device type was not significant ($\beta = -1.11$, $SE = 1.89$, $p = .557$), indicating no reliable difference in mental demand between mobile and desktop users. The interaction between task difficulty and device type was also non-significant ($\beta = 1.65$, $SE = 1.12$, $p = .141$).
These results support the descriptive findings, confirming that harder tasks reliably increased perceived mental workload, independent of the device used. \autoref{fig:tlx_plots} illustrates this trend for all NASA-TLX subscales.

\renewcommand{\thefootnote}{\fnsymbol{footnote}}
\footnotetext[1]{%
$R^2_{\text{marginal}}$ = variance explained by fixed effects;
$R^2_{\text{conditional}}$ = variance explained by both fixed and random effects;
$\beta$ = estimated regression coefficient (effect size); 
$SE$ = standard error; 
CI = confidence interval.
}

\subsubsection{Mental Effort}
We also asked participants to rate the mental effort required to write and refine the meal plan prompt every interaction.
We found a significant main effect of task difficulty (p < 0.05) on mental effort, indicating that more demand was imposed by the hard tasks ($M = 11.97, SD = 5.83$) than the easy tasks ($M = 8.08, SD = 5.42$). No significant main effect was found for device type, and there was no significant interaction effect. The specific means were: desktop-easy ($M = 7.24, SD = 6.09$), desktop-hard ($M = 11.82, SD = 6.05$), mobile-easy ($M = 9.14, SD = 4.47$), and mobile-hard ($M = 12.13, SD = 5.76$).

\subsubsection{LLM Output Usefulness}
We also asked them to rate each LLM output by it's usefulness for achieving the task.
There was a significant main effect of task difficulty on output usefulness (p < 0.001), with the response usefulness during easy tasks  ($M = 13.71, SD = 5.22$) being rated significantly higher than during hard tasks ($M = 8.83, SD = 5.26$). There was no significant main effect of device type and no significant interaction effect. The means for the specific conditions were: desktop-easy ($M = 14.15, SD = 4.67$), desktop-hard ($M = 8.96, SD = 5.34$), mobile-easy ($M = 13.27, SD = 5.82$), and mobile-hard ($M = 8.70, SD = 5.33$).

\subsubsection{Difficulty Refining Output}
Regarding refinement difficulty, the analysis revealed a significant main effect for task difficulty (p < 0.05). Overall, refining the LLM output for the hard task ($M = 11.85, SD = 5.49$) was rated as more difficult than for the easy task ($M = 8.18, SD = 5.45$). There was no significant main effect for device type and no significant interaction effect. The condition means were: desktop-easy ($M = 6.41, SD = 5.53$), desktop-hard ($M = 11.76, SD = 5.32$), mobile-easy ($M = 10.38, SD = 4.66$), and mobile-hard ($M = 11.95, SD = 5.81$).

\subsubsection{Confidence}
Subjective confidence ratings, assessed using a 21-point scale ranging from 0 (very low) to 20 (very high), were collected to evaluate users’ perceived success in meeting task requirements. Confidence was substantially higher in easy tasks (Desktop–Easy: $M = 16.56$, $SD = 6.34$; Mobile–Easy: $M = 17.11$, $SD = 4.04$) than in hard tasks (Desktop–Hard: $M = 8.06$, $SD = 7.57$; Mobile–Hard: $M = 8.39$, $SD = 7.77$).
There was a significant main effect of task difficulty (p < 0.001) on confidence ratings. Participants reported significantly higher confidence after the easy tasks compared to the hard tasks. There was no significant main effect for device type and no significant interaction effect.

\subsection{Predicting LLM Usefulness from Keystrokes}

To assess whether behavioral typing metrics could predict participants' perceived usefulness of AI output, we fitted three distinct models: a linear mixed-effects model, a principal component regression, and a non-linear random forest model. These models incorporated five core typing metrics: Words Per Prompt, Keystrokes, Backspace Usage Ratio, Inter-Key Interval, and Pause Count.

In the linear mixed-effects model, none of the individual typing metrics significantly predicted feedback usefulness. Fixed effects accounted for only 1\% of the variance ($R^2 = 0.01$), while total model variance explained was 25\% ($R^2_{\text{conditional}} = 0.25$). All predictors (e.g., WPP: $p = .903$, Keystrokes: $p = .676$, Pauses: $p = .759$) failed to reach statistical significance.
Principal component regression using the top two components (PC1, PC2) yielded similar results. PC1 showed a marginal trend toward significance ($\beta = -0.35$, $p = .052$), but did not meet conventional thresholds. Overall, the model retained low explanatory power ($R^2 = 0.01$).
Finally, a random forest model trained with 5-fold cross-validation produced minimal predictive performance, with a best-case $R^2$ of 0.005. This suggests that the relationship between typing behavior and perceived usefulness is weak or non-existent in this dataset.
These findings collectively indicate that the typing behavior metrics examined in this study do not reliably predict participants' ratings of the LLM's response usefulness.

\begin{table}[hb]
\centering
\footnotesize
\caption{Summary of modeling results for predicting AI output usefulness from typing behavior.
\add{$R^2_{\text{marg}}$ denotes variance explained by fixed effects, and $R^2_{\text{cond}}$ variance explained by the full model (fixed and random effects). For the random forest, we report mean cross-validated $R^2$.}}
\label{tab:utility-model-summary}
\begin{tabular}{lcccc}
\toprule
\textbf{Model} & \textbf{$R^2_{\text{marg}}$} & \textbf{$R^2_{\text{cond}}$} & \textbf{Best Predictors} & \textbf{Significant?} \\
\midrule
LMM & 0.01 & 0.25 & None & No \\
PCR (PC1, PC2) & 0.01 & 0.25 & PC1 (borderline) & No \\
Random Forest & \multicolumn{2}{c}{Mean $R^2$: 0.005} & None & No \\
\bottomrule
\end{tabular}
\end{table}

\section{Discussion}

\subsection{\add{Summary of Findings}}

This study explored how typing behavior reflects users' perceived utility of LLM output and cognitive effort during prompting LLMs. Keystroke data were collected across different task-difficulty conditions and device types.

\subsubsection{\add{Typing Behavior Across Task Complexity and Device Type (RQ1, RQ2)}}

\paragraph{Keystrokes capture cognitive effort during human-LLM interaction.} Overall, task difficulty had a clear and consistent impact. In harder tasks, participants wrote longer and more frequent prompts, used more keystrokes, paused more often, and typed more slowly. 
\add{These findings are in line with prior work, which similarly reported lower typing speed and higher pause frequency during high-demand writing tasks~\cite{oliveira2020writing}.}\delete{The increased keystroke count should be interpreted cautiously, as the complex task inherently demanded more input, with task requirements nearly three times as long as those of the simpler task in terms of character count.} Additionally, participants reported significantly higher levels of mental demand, frustration, and effort. These trends suggest that harder tasks require deeper reflection and engagement, even with the aid of LLMs.
\add{While it is inherently expected that a more complex task, with more requirements, would necessitate a higher volume of interactions and keystrokes, our findings suggest that the cognitive load is not merely a byproduct of text length. If the increased interaction were purely mechanical, we would expect typing speed to remain stable as volume increased. Instead, the observed decrease in typing speed and the increase in pause frequency indicate that participants were engaged in more frequent "think-then-type" cycles during the hard task. 
This suggests that the hard task successfully triggered a loop of cognitive evaluation: users had to parse the LLM's output, compare it against multiple constraints, and strategize a refinement. The keystroke data, therefore, captures the friction of this mental processing rather than just the physical labor of inputting the task requirements.}

\delete{The success in reflecting cognitive effort across conditions suggests promising avenues for adaptive system design. By leveraging keystroke-based indicators of user effort, future platforms could dynamically adjust their behavior to better support users in real time. Several layers of adaptivity could be explored. On the interface level, systems could adjust the interface to reduce cognitive strain. On the model level, the choice of LLM or its response length and verbosity could be tailored to the user’s cognitive state. Finally, semantic-level adaptations could refine the content or complexity of LLM responses, ensuring that they align more closely with the user’s needs and capacity in the moment. Such adaptive mechanisms would mark a significant step toward more intelligent, user-aware human-AI collaboration.}

\paragraph{Keystrokes are equally expressive of cognitive effort across mobile and desktop.}
In contrast, the type of device used (mobile vs. desktop) did not produce strong main effects. Participants typed more slowly on mobile devices, which goes in hand with known previous findings\add{~\cite{barrett1994performance, palin2019people}}, but other differences were limited. For instance, word count and the perceived mental demand remained consistent across both device types. It is worth noting that the interaction in the mobile and desktop condition was nearly identical in terms of interface design, system responsiveness, and overall smoothness. The main difference was the typing method (physical keyboard versus touchscreen) and the size of the input interface. However, given the relatively young and digitally literate participant pool, it is likely that most users were sufficiently comfortable with both input modalities, resulting in minimal observable impact on typing behavior or cognitive workload.

\paragraph{Mobile subtly modulated typing behavior under high difficulty.}
When we looked at the interaction between task difficulty and device type, the effects were generally weak. One exception was a decrease in keystrokes for hard tasks on mobile compared to desktop, despite the overall increase due to task complexity. This may reflect a tendency among mobile users to shorten or simplify their responses under more constrained input conditions due to smaller key and screen sizes.
\add{This is a critical finding for the relationship between task complexity and typing volume. If keystroke count were solely a function of task requirements, we would see a uniform increase across both devices. However, we see evidence that typing behavior is actively modulated by cognitive load and interface constraints, not merely by the number of task requirements.}

\paragraph{Backspace use is unaffected by device type and task difficulty.}
While keystroke count, words per prompt, pause count, and inter-key interval were all significantly impacted by task difficulty, backspace use remained consistent across all conditions. It did not change with task difficulty or device. This suggests that editing habits are resilient to variations in task complexity or device constraints.

\subsubsection{\add{Predicting Usefulness from Keystrokes (RQ3)}}
\paragraph{
Keystrokes do not capture perceived LLM output usefulness.}
Perceived usefulness of LLM outputs declined significantly for hard tasks despite increased user effort, \delete{indicating}\add{a pattern consistent with} diminishing returns from additional interaction and refinement under higher task complexity\add{, though no causal relationship can be inferred}. Our results underscore the task-complexity-polarization phenomenon in human-AI collaboration, which has been shown in previous works, postulating that easy tasks remain easy, but hard tasks stay hard despite the assistance of AI~\cite{simkute2025ironies}.

\delete{However, typing behavior}\add{The measured keystroke metrics} could not predict the perceived usefulness of AI responses. In our models, none of the keystroke metrics correlated with participant ratings of response usefulness. At the same time, our dataset showed sufficient variation in perceived utility, with participant assessments ranging from highly useful to not useful at all. This suggests that the absence of correlation is not due to a lack of variance in the target variable. A more likely explanation is that usefulness is shaped primarily by the semantic content of the prompt \delete{rather than by how it was typed}\add{or keystroke metrics not captured within this study}. 
\add{If the LLM fails to sufficiently satisfy the task demands, users may engage in more iterative prompt refinement, as evidenced by longer pauses between bursts of writing, reflecting planning and reformulation. Another indicator of diminishing returns might be a high lexical overlap but limited semantic progression, increasing edit distance between successive prompts without corresponding improvements in output. These patterns might suggest that users invest additional effort in attempting to steer the model, yet receive limited benefit.}

\add{Overall,} while keystroke dynamics offer a window into user effort and cognitive load, they appear less effective for capturing higher-level subjective \delete{evaluations}\add{judgements} such as usefulness. More fine-grained metrics \add{and semantic text analysis} could offer additional insights. Similarly, combining keystroke data with other behavioral signals such as mouse movement or scrolling patterns might help model perceived usefulness more accurately. 

\hfill\\
In summary, typing behavior captured real-time user effort and showed some variation across devices. 
\add{These findings suggest that keystroke dynamics can serve as an accessible proxy for mental demand during human-AI collaboration and contribute to the growing HCI literature on implicit user state detection~\cite{chiossi2024evaluating, schuetz2025eye2, shaikh2025general, funk2025eye}.}
However, a comprehensive understanding of certain user perceptions, such as collaboration success, may require incorporating additional signals within and beyond keystroke dynamics.

\subsection{\add{Designing Adaptive Human-LLM Interactions Based on Keystrokes}}

\subsubsection{\add{Cognitive Load Adaptive Human-AI Interactions}}
\add{The success in reflecting cognitive effort across conditions suggests promising avenues for adaptive system design. By leveraging keystroke-based indicators of user effort, future platforms could dynamically adjust their behavior to better support users in real time. Several layers of adaptivity could be explored. On the interface level, systems could adjust the interface to reduce cognitive strain. On the model level, the choice of LLM or its response length and verbosity could be tailored to the user’s cognitive state. Finally, semantic-level prompt adaptations could refine the content or complexity of LLM responses, ensuring they align more closely with the user’s needs and capacity in the moment.}

\add{Building on prior work on adaptive prompting interfaces~\cite{dang2023prompting}, such signals could also guide when to provide assistance. For instance, systems might withhold prompt-completion suggestions under low cognitive load while offering completions as effort increases. 
Alternatively, keystroke dynamics could serve as real-time signals that inform adaptive task allocation strategies, as proposed by \citet{hemmer2023human}, by dynamically shifting responsibility between human and AI based on the most suitable actor.
Importantly, these signals may also support meta-level interventions. Systems could detect when continued LLM interaction yields diminishing returns and notify users accordingly, encouraging more deliberate and context-sensitive AI usage. This aligns with prior findings that users do not always make optimal decisions about when to rely on AI, even when performance information is available~\cite{qiao2025overreliance}.}

\subsubsection{\add{Balancing Cognitive Offloading and Critical Thinking}}
\add{While cognitive-load adaptive systems hold promise, future HCI research must balance the goal of reducing cognitive load with the risk of user deskilling~\cite{shukla2025deskilling}. Lower cognitive load is not always desirable if it removes opportunities for reflection and skill acquisition during AI-assisted tasks, as prior work shows that users may reject low-effort recommendation systems in favor of approaches that keep them cognitively engaged~\cite{reicherts2025myself}.
Interfaces should therefore be designed not only to optimize efficiency, but to deliberately preserve moments of cognitive engagement that are critical for reflection, verification, and independent reasoning. This aligns with prior findings that increased confidence in AI can reduce critical thinking effort, suggesting that overly assistive interfaces may inadvertently encourage cognitive offloading rather than active stewardship~\cite{reicherts2025myself, lee2025critical}. Consequently, adaptive systems should incorporate friction, transparency, and user-controlled levels of assistance to maintain critical thinking and skill development.}

\subsubsection{Privacy-Preserving Keystroke Data Processing}
While keystroke dynamics offer a powerful method for evaluating user state, logging keystroke data raises significant privacy concerns. Keystroke patterns can be unique enough to serve as a biometric identifier~\cite{martins2025keystroke} and could be used for the inference of sensitive traits beyond cognitive load, such as emotional states~\cite{qi2021emotion, trojahn2013emotion, epp2011identifying} or health conditions~\cite{alfalahi2022diagnostic}. In our study, data was processed locally and anonymized. However, real-world deployment must prioritize privacy-preserving architectures, using edge-based processing or federated learning approaches to refine models without centralizing sensitive biometric data.

\subsection{Limitations \& Future Work}
We acknowledge several limitations of our work. The participant pool was relatively homogeneous, with most participants being university students aged between 19 and 34 years. This affects how findings apply to broader populations. 
Keystroke data was limited to key press events only. We did not record key release timestamps, which restricted the analysis of metrics such as dwell time and flight time.

\add{Another limitation relates to the analysis of a single task (meal planning), which limits the generalizability of the findings. Future work should repeat this study across multiple task domains. Furthermore, we utilized a single local LLM instance. Replicating the study with various state-of-the-art LLMs, as well as comparing the results to non-LLM chatbots, would help isolate the specific impact of human-agent interaction dynamics.}

This work represents an initial step toward understanding interaction-level signals in human-AI collaboration. Potential next steps could include semantic analysis of user inputs. Whereas keystroke patterns capture interaction effort, the prompts themselves can reveal aspects of user state, such as emotional tone or cognitive complexity. Metrics such as lexical diversity, syntactic structure, and sentiment may help explain why some responses are perceived as more useful than others. A multi-layered prompt analysis combining behavioral and semantic signals may provide a more complete picture of perceived utility in LLM-assisted tasks.
In addition, future studies could extend the current logging system to include key-up events, enabling the calculation of dwell and flight times.
To broaden the scope of interaction data, subsequent research may also incorporate multimodal interaction signals, such as mouse \cite{lim201detecting} and eye tracking \cite{shi2024crtypist}.

\section{Conclusion}
This work explored whether keystroke dynamics can serve as a real-time indicator of user effort and perceived utility during interactions with large language models. We conducted a controlled user study manipulating task difficulty and device type to examine their influence on typing behavior and subjective experience. 
Our findings showed that keystroke dynamics can reliably reflect cognitive effort during prompting across mobile and desktop devices. While keystroke dynamics did not predict perceived usefulness of LLM output, we believe that combining typing metrics with other biosignals or sentiment analysis holds promise.
Overall, our work demonstrates the potential of keystroke data as a low-cost, non-intrusive method for designing adaptive AI systems that respond to users' cognitive states in real time. Finally, as LLMs continue to evolve and expand in use, understanding how users engage with them remains a pressing challenge. We see this work as an important step toward designing more human-centered AI systems.

\vspace*{-2.5pt}
\section{Data Availability}
The collected data and analysis scripts are available on Open Science
Framework (\href{https://osf.io/dsnqf}{https://osf.io/dsnqf}).

\vspace*{-2.5pt}
\begin{acks}
This work was conducted as part of the AI-Twin project, which is funded by the European Research Council (ERC-2024-ADG) as part of the European Union’s Horizon 2020 research and innovation program (grant agreement no. 101200584).
\end{acks}

\bibliographystyle{ACM-Reference-Format}
\bibliography{bibliography}

\newpage

\appendix
\section{Task Instructions}
\label{appendix:task-instructions}

How it works: Interact with our LLM to create a 7-day meal plan based on the list of requirements. After each LLM response, you will be asked to evaluate it through a short form. Keep interacting with the LLM and improving the
meal plan until you are fully satisfied with the LLM's response. When you're
happy with it, click “Done” to end the task. You'll be asked to fill out one last
evaluation form before submitting your work.

\subsection{Easy Task}

Your task is to work with our language model to create a 7-day meal plan
that meets the following requirements:
\begin{enumerate}
    \item 7-day meal plan
    \begin{itemize}
        \item 5 days: 3 meals + 1 snack per day
        \item 2 days: 2 meals only per day
    \end{itemize}
    \item No processed food
    \item Dairy-free
\end{enumerate}

\subsection{Hard Task}

Your task is to work with our language model to create a 7-day meal plan
that meets the following requirements:
\begin{enumerate}
    \item 7-day meal plan
    \begin{itemize}
        \item 5 days: 3 meals + 1 snack per day
        \item 2 days: 2 meals only per day
    \end{itemize}
    \item No repeated dishes
    \item No repetition of main ingredients on 2 consecutive days
    \item Even distribution of calories: ~ 2000 per day
    \item High in protein, low in carb meal plan
    \item Show the percentage of carbs, fats, and proteins for every day (sum must add up to 100\%)
    \item Dairy-free
\end{enumerate}

\end{document}